\documentclass[trackchanges, twocolumn]{aastex701}
\usepackage{amsmath}

\newcommand{\Mstar}{M_\star }
\newcommand{\SFR}{{\rm SFR} }

\newcommand{\OH}{12+\log_{10}(\rm{O/H})}
\newcommand{\FeH}{12+\log_{10}(\rm{Fe/H})}

\shorttitle{Not So Heavy Metal I: the Rate of High-z SLSNe}
\shortauthors{Margalit et al.}

\begin{document}

\title{Not So Heavy Metal I: 
Low Metallicity Enhances the Rate of high-redshift Superluminous Supernovae
}

\newcommand{\UMN}{\affiliation{School of Physics and Astronomy, University of Minnesota, Minneapolis, MN 55455, USA}}

\newcommand{\NU}{\affiliation{Department of Physics and Astronomy, Northwestern University, Evanston, IL, USA}}

\newcommand{\CIERA}{\affiliation{Center for Interdisciplinary Exploration and Research in Astrophysics, Northwestern University, Evanston, IL, USA}}

\newcommand{\SkAI}{\affiliation{NSF-Simons AI Institute for the Sky (SkAI), 172 E. Chestnut St., Chicago, IL 60611, USA}}

\newcommand{\Weizmann}{\affiliation{Department of Particle Physics and Astrophysics, Weizmann Institute of Science, 234 Herzl St, 76100 Rehovot, Israel}}
\author[0000-0001-8405-2649]{Ben Margalit}
\UMN
\email{margalit@umn.edu}

\author[0000-0001-6369-1636]{Allison L. Strom}
\NU
\CIERA
\email{allison.strom@northwestern.edu}

\author[0009-0004-9687-3275]{Cristina Andrade}
\UMN 
\email{andra104@umn.edu}

\author[0000-0002-8262-2924]{Michael W. Coughlin}
\UMN 
\email{cough052@umn.edu}

\author[0000-0001-9515-478X]{Adam A. Miller}
\NU
\CIERA
\email{amiller@northwestern.edu}

\author[0000-0001-6797-1889]{Steve Schulze}
\Weizmann
\email{steve.schulze@weizmann.ac.il}

\author[0009-0009-1590-2318]{Ved G. Shah}
\NU
\CIERA
\email{VedShah2029@u.northwestern.edu}

\correspondingauthor{Ben Margalit}
\email{margalit@umn.edu}

\begin{abstract}
Type-I superluminous supernovae (SLSNe) are extremely luminous stellar explosions originating from the core collapse of massive stars. Unlike normal core-collapse supernovae, however, SLSNe occur almost exclusively in metal-poor environments. As a result, their volumetric rate should not simply trace the cosmic star-formation history, 
but instead evolve with the metallicity distribution of galaxies across cosmic time. In this work, we introduce a new theoretical framework for modeling the volumetric rate of astrophysical sources as a function of redshift and metallicity. 
Our formalism combines metallicity scaling relations, stellar-mass distribution functions, and metallicity-dependent production efficiencies. 
Importantly, it also distinguishes between metallicity measures based on O/H and Fe/H abundances.
Applying our new formalism to SLSNe, we predict volumetric
rates that are nearly an order of magnitude higher at $z \gtrsim 2$ compared to models in which the rate traces the cosmic star-formation rate alone. 
This has direct implications 
on the detection prospects of high-$z$ SLSNe with the Vera C. Rubin Observatory and other time-domain surveys.
\end{abstract}

\keywords{
Supernovae (1668), Metallicity (1031), Galaxies (573), Star formation (1569)
}

\section{Introduction}

As we enter the era of extremely wide and deep time-domain surveys, best exemplified by the forthcoming surveys from the Vera C.~Rubin Observatory \citep{Ivezic19} and the Nancy Grace Roman Space Telescope \citep{Akeson19_roman}, there will be a dramatic increase in the discovery of supernovae (SNe) at redshifts that are inaccessible to current $\sim$1\,m class telescope surveys. 
Superluminous supernovae (SLSNe; 
see \citealt{Gal-Yam12,Gal-yam19,Howell17,Nicholl21} for reviews) 
will be especially interesting in this respect because their high intrinsic luminosity will enable us to probe these events at relatively high redshifts \citep[e.g.,][]{Villar+18}.

Despite intensive study, and a growing catalog that includes $>$100 examples \citep{Nicholl+17b,Chen+23,Gomez24,Aamer+25}, the predominant power source for hydrogen-poor `Type-I' SLSNe (hereafter simply referred to as SLSNe) remains unexplained. 
One possibility is that these rare explosions are the result of pair-instability SNe \citep[PISNe;][]{barkat67}, which can produce the several solar mass of freshly synthesized $^{56}$Ni required to explain the peak luminosity of these transients. This interpretation is not favored, however, as the photometric evolution of SLSNe proceeds on timescales that are too short to be consistent with this picture \citep[e.g.,][]{quimby05ap,Gomez24}, with only a few exceptions \citep[e.g.,][]{Schulze24}. 
Instead, it is commonly assumed that SLSNe are powered by a ``central engine,'' typically a millisecond magnetar that injects its rotational energy into the expanding SN ejecta (e.g., \citealt{kasen10,Woosley10,Metzger+15,Margalit+18,Vurm21}), however, fall-back accretion has also been suggested as a possible engine (\citealt{Dexter&Kasen13}; though see \citealt{Moriya+18}).
While the light curves of many SLSNe are well-matched by the magnetar model \cite[e.g.,][]{Nicholl+17b,Chen+23,Gomez24}, there remain many degrees of freedom in these models.
An alternative model that is favored for Type-II SLSNe, but is also plausible for some Type-I SLSNe, suggests that collisions between the SN ejecta and ambient circumstellar medium (CSM) could explain the high luminosities of SLSNe (e.g., \citealt{Chevalier11,Ginzburg&Balberg12,Chatzopoulos+12,Sorokina+16}; see \citealt{Moriya_Sorokina_Chevalier_2018} for a review of SLSNe powering mechanisms).

An interesting clue to the origin and physics of SLSNe comes from their birth environment. 
SLSNe are observed to occur primarily in low-mass dwarf galaxies. Because such galaxies correlate with lower-metallicity environments \citep[e.g.,][]{Tremonti+04} it has been suggested that SLSN production can only occur at low metallicities \citep[e.g.,][]{Neill11,Lunnan+14,Perley16,Chen17a,schulze2018,Schulze21}. 
While these dwarf galaxies are actively forming stars, they are a minor contributor to the total star formation in the local Universe \citep{Gilbank2010_local_SFRD, Li2011_SFRD, Murrell2025_SFRD}. 
This implies that, unlike standard CCSNe, SLSNe are \textit{biased} tracers of star formation. 

Despite this observation, the SLSN rate has typically been assumed to track the cosmic star-formation rate density \citep[SFRD; e.g.,][]{Villar+18,Kessler19}. 
This assumption is motivated by the fact that CCSNe originate from short-lived massive stars whose birth rate is determined by the local star formation rate (SFR). Indeed, as noted above, standard Type-II CCSNe do seem to follow this expected trend \citep[e.g.,][]{Horiuchi11}. 
However, the apparent metallicity aversion of SLSNe means that the cosmic evolution of metallicity must also be taken into account when extrapolating the local SLSN rate to high redshifts.

In this {\it Letter}, we explicitly model this scenario and calculate the predicted SLSN rate evolution $\mathcal{R}(z)$ accounting for the metallicity dependence of these events.
We begin by introducing our formalism in \S\ref{sec:formalism}. Importantly, our novel formalism distinguishes between metallicity measures based on alpha-elements (typically based on O/H abundance measurements) and iron-group elements (\S\ref{sec:Fe-formalism}). While iron is likely more important for setting the physical conditions and stellar properties that determine SLSN-production, previous consideration of metallicity effects has typically been restricted to O/H measures.
In \S\ref{sec:SLSN_application} we apply our formalism to SLSNe and explicitly calculate the volumetric rate of SLSNe as a function of redshift. We discuss our results in \S\ref{sec:discussion} and conclude in \S\ref{sec:conclusions}.

Throughout the rest of this paper we adopt solar abundances from \citet{Asplund09}, namely $12+\log_{10}(\textrm{O/H})_\odot = 8.69$ and $12+\log_{10}(\textrm{Fe/H})_\odot = 7.50$. As a shorthand, we will refer to metallicities 
measured based on the abundance of element ${\rm X}$
as $\log_{10}(Z_{\rm X}/Z_\odot) = \log_{10}(\rm{X/H})-\log_{10}(\rm{X/H})_\odot$.

\section{Formalism: metallicity-dependent rate calculation}
\label{sec:formalism}

We assume that the formation rate of some astrophysical source (in our case, SLSNe) is dependent on the metallicity $Z$ with which the progenitor is born. For SLSNe, the data points towards a strong suppression above some cutoff metallicity $Z_{\rm max}$, which we model as a sigmoid with the following functional form,
\begin{equation}
\label{eq:eta}
    \eta \left( Z ; Z_{\rm max} \right) = 1 - \frac{1}{ 1 + \left( Z / Z_{\rm max} \right)^{-\beta} }
    .
\end{equation}
Equation~(\ref{eq:eta}) can be thought of as a smoothed step-function that returns $\eta \approx 1$ when $Z \ll Z_{\rm max}$ and $\eta \approx 0$ at high metallicity $Z \gg Z_{\rm max}$. The parameter $\beta$ controls the width of the transition between these two regimes. 
In our subsequent calculations we adopt $\beta = 50$. While this specific choice is ad hoc, we note that a large value of $\beta$ is driven by the limited dynamic range in $Z$.  
For example, the average oxygen abundance of a galaxy only varies from $\OH \approx 8.6$ 
to $\OH \approx 9.1$ 
as the galaxy stellar mass ($M_\star$) jumps two orders of magnitude from $\Mstar = 10^9 M_\odot$ to $\Mstar = 10^{11} M_\odot$ \citep{Tremonti+04}.
This translates into a factor $\sim 3$ change in metallicity, from
oxygen-inferred metallicity 
$Z_{\rm O} \approx 0.8\,Z_\odot$ to $\approx2.6\,Z_\odot$.
For $\beta = 50$ the function $\eta(Z)$ is effectively a step-function, so the results do not depend on the precise value of $\beta$ so long as $\beta \gg 1$. The value of $Z_{\rm max}$ plays a more critical role in our calculations, and we discuss choices for this parameter in \S\ref{sec:metallicity_cutoff}.
Finally, note that the formalism described in this section applies regardless of the functional form of $\eta(Z)$, which could be qualitatively different for other astrophysical sources.

We further assume that the metallicity suppression discussed above is the sole determinant of SLSN production efficiency. That is, we assume that the number of SLSNe produced is proportional to the SFR (itself a proxy for the number of massive stars being born) that occurs in regions where the metallicity is below the threshold $Z_{\rm max}$. This implies that the SLSN rate is 
\begin{equation}
\label{eq:Rate}
    \mathcal{R}(z) \propto \Psi(z) f(z;Z_{\rm max})
\end{equation}
where $\Psi(z)$ is the cosmic SFRD \citep[e.g.,][]{Madau&Dickinson14} and the function $f(z;Z_{\rm max})$ can be interpreted as the fraction of SFR that occurs at metallicities $< Z_{\rm max}$, at some redshift $z$. The problem of calculating the SLSN rate evolution is therefore limited to a calculation of the function $f(z;Z_{\rm max})$.

This can be done provided a multi-dimensional galaxy distribution function $\rho \left( \Mstar, \SFR, Z \vert z \right)$ that describes the relative abundance of galaxies\footnote{
Note that this approach implicitly assumes that global galaxy properties (e.g., metallicity and total SFR) are indicative of local properties within these galaxies. This is a common assumption and caveat in studies of transient host-galaxies. 
For SLSNe, the relative compactness of typical SLSNe hosts \citep[e.g.,][]{Lunnan+15} mitigates this concern to some extent.
} 
with stellar mass $\Mstar$, star-formation rate SFR, and metallicity $Z$ at any redshift $z$.
The fraction of star formation that occurs in galaxies where the metallicity is low enough to potentially harbor SLSNe can then be calculated as
\begin{equation}
\label{eq:f_general}
    f(z) 
    = \frac{ 
    \int \SFR\, \eta \left( Z \right) \rho \left( \Mstar, \SFR, Z \vert z \right) \, d\Mstar d\SFR dZ 
    }{  
    \int \SFR\, \rho \left( \Mstar, \SFR, Z \vert z \right) \, d\Mstar d\SFR dZ
    }
    .
\end{equation}
Although a galaxy probability density function of the form $\rho \left( \Mstar,\SFR \vert z \right)$ has recently been put forward by \cite{Leja+22}, to our knowledge there is no generalized function $\rho \left( \Mstar, \SFR, Z \vert z \right)$ that explicitly includes metallicity, an essential ingredient in our calculations.
We therefore simplify Equation~(\ref{eq:f_general}) by employing various scaling relations, and re-write $f(z)$ in terms of more readily-available quantities.

Most importantly, we adopt a relation that connects $Z$ to both the stellar-mass and SFR of a given galaxy.
Many studies have shown that the metallicity of a galaxy, typically quantified by the gas-phase oxygen abundance $\OH$, is strongly correlated with the stellar mass $\Mstar$ of the galaxy \citep[e.g.,][]{Tremonti+04}. Further studies have also argued that the scatter around this mass-metallicity relation (MZR) can be reduced by accounting for the galaxy's SFR, in what is known as the fundamental metallicity relation \citep[FMR; e.g.,][]{mannucci2010}.
Although there remains significant debate about the the redshift invariance of the FMR and the exact relationship between metallicity and star formation at fixed stellar mass at $z \gtrsim 1$ \citep[e.g.,][]{korhonencuestas2025}, we proceed with this more-general formalism. It would be straightforward to replace our adopted FMR with a different functional choice (which could be redshift-dependent) or, alternatively, with an MZR.
In our present work,
we adopt the FMR given by Equation~(5) from \cite{mannucci2010}.
This provides an estimate of a galaxy's oxygen abundance as a function of its stellar mass and SFR,
\begin{equation}
    \label{eq:FMR}
    12 + \log_{10}({\rm O/H})
    = 
    \begin{cases}
        8.90 + 0.47 x &;\, x<17/47
        \\
        9.07 &;\, {\rm else}
    \end{cases}
\end{equation}
where
\begin{equation}
    x = \log_{10}(M_\star) - 0.32 \times \log_{10}({\rm SFR}) - 10
    .
\end{equation}

Employing an FMR such as the one described above allows us to convert between $Z \leftrightarrow (M_\star,{\rm SFR})$ and
write the metallicity efficiency function Equation~(\ref{eq:eta}) as $\eta(\Mstar,\SFR)$.
We can then get rid of $Z$ as an independent variable and simplify Equation~(\ref{eq:f_general}).
In place of the galaxy probability distribution function $\rho$, we recast Equation~(\ref{eq:f_general}) in terms of the stellar mass function $\Phi(\Mstar \vert z)$ and the distribution of galaxies around the star-forming main sequence (SFMS), $\varphi(\SFR \vert \Mstar,z)$.
With these assumptions, we rewrite Equation~(\ref{eq:f_general}) as
\begin{equation}
    \label{eq:f_final}
    f(z) 
    = \frac{ 
    \int \SFR\, \eta \left( Z \right) \varphi(\SFR \vert \Mstar,z) \Phi \left( \Mstar \vert z \right) \, d\Mstar d\SFR
    }{  
    \int \SFR\, \varphi(\SFR \vert \Mstar,z) \Phi \left( \Mstar \vert z \right) \, d\Mstar d\SFR
    }
    .
\end{equation}

In our present work, we adopt a stellar mass function from \cite{Leja+20} 
and assume that $\varphi(\SFR \vert \Mstar,z) \sim \mathcal{N}( \log_{10}\SFR \vert \mu,\sigma)$, such that galaxies are distributed log-normally around the SFMS with standard deviation $\sigma = 0.3\,{\rm dex}$ and mean $\mu(\Mstar,z)$ set by the \cite{Leja+22} SFMS `ridge' model.
Note that the denominator in Equation~(\ref{eq:f_final}) can be simplified since
$\int \SFR\, \varphi(\SFR \vert \Mstar,z) \, d\SFR = 
10^{\mu(\Mstar,z)}$ 
is the mean SFR, as determined by the SFMS.
A similar simplification cannot, in general, be performed for the numerator because the metallicity $Z$ incorporates some dependence on SFR (see Equation~\ref{eq:FMR}).

\subsection{Accounting for Fe-group Metallicity}
\label{sec:Fe-formalism}

The formalism above is quite general. A key ingredient that is required in order to calculate $f(z)$ following Equation~(\ref{eq:f_final}) is an MZR or FMR relating metallicity $Z$ to other galaxy properties (namely $M_\star$ and SFR). However, an important question that arises here is: {\it what do we mean by metallicity?}
In most practical applications, star-forming galaxy metallicity is quantified via the abundance of oxygen. The metallicity encoded in an FMR (e.g., Equation~\ref{eq:FMR}) should therefore be interpreted as $Z_{\rm O}$. However, as we argue below, this may not be a particularly physically-motivated choice when discussing the formation efficiency of many astrophysical sources.

It is well known the the abundance of iron-group elements plays an important role in stellar evolution, influences binarity, and possibly affects the initial-mass function \citep[e.g.,][]{Moe+19,Sana+25}. By comparison, the abundance of oxygen---in and of itself---is not thought to play an important role in these processes. 
This suggests that the underlying driver of metallicity effects on the birth rate of astrophysical sources such as SLSNe is likely associated with the abundance of iron-group elements. By contrast, most discussion of metallicity effects on such sources uses metallicity based on gas-phase oxygen abundances, which is typically easier to measure. A nuanced distinction between iron-based and oxygen-based metallicities is therefore missing in most previous work. This distinction is especially important because 
Fe and O do not evolve in lockstep with one another but instead depend sensitively on the star-formation history of a galaxy.

We therefore proceed with what we consider to be a more physically-motivated choice---that the metallicity dependence of SLSN production is related to the abundance of iron, so that $\eta = \eta(Z_{\rm Fe})$.
In order to model this, we need to relate Fe/H to $M_\star$ and SFR. This requires a conversion between O/H (the typical constrained abundance in galaxy metallicity-relations) and Fe/H.
This conversion is non-trivial and will change over the course of a galaxy's lifetime since iron and oxygen are produced through different channels: oxygen is produced in massive stars and released back to the ISM via CCSNe; while iron is also produced in CCSNe, it is additionally fused in Type Ia SNe. 
Comprehensively modeling these competing factors is beyond the scope of this paper, but this issue has recently been investigated by \cite{Chruslinska+24,Chruslinska+26}. In the following, we adopt the `mixed Fe-enrichment' model of \cite{Chruslinska+26} for our fiducial estimates. This model specifies the relative oxygen to iron abundance $\log_{10}({\rm O}/{\rm Fe})[{\rm sSFR}]$ in terms of a galaxy's specific SFR, ${\rm sSFR} = \SFR / \Mstar$.
The iron abundance is then determined via
\begin{equation}
    \label{eq:FeH}
    \log_{10}({\rm Fe/H}) = \log_{10}({\rm O/H}) - \log_{10}({\rm O/Fe})[{\rm sSFR}]
    ,
\end{equation}
where the oxygen abundance O/H can be specified using a standard FMR.
Combining Equations~(\ref{eq:FMR},\ref{eq:FeH}) therefore allows us to estimate the iron-based metallicity of a galaxy as a function of galaxy mass and SFR, $Z_{\rm Fe}(M_\star,{\rm SFR})$.
We use this relation to calculate $f(z)$ via Equation~(\ref{eq:f_final}) for our fiducial model.
As a rough estimate of the uncertainty induced by the $({\rm O}/{\rm Fe})$ conversion, we bracket our fiducial results between two extremes derived by adopting the `slow' and `fast' Fe-enrichment models of \cite{Chruslinska+26} (see their Figure~1). These describe competing scenarios for the delay-time and yield of iron-enrichment due to SNe Ia.
Finally, we also consider a reference O-dependent model in which $Z = Z_{\rm O}$, and the metallicity threshold is assumed to depend on the oxygen abundance rather than iron. This model mainly serves for comparison purposes.

\section{Application to SLSNe}
\label{sec:SLSN_application}

\subsection{Metallicity Cutoff}
\label{sec:metallicity_cutoff}

\begin{figure}
    \centering
    \includegraphics[width=0.99\linewidth]{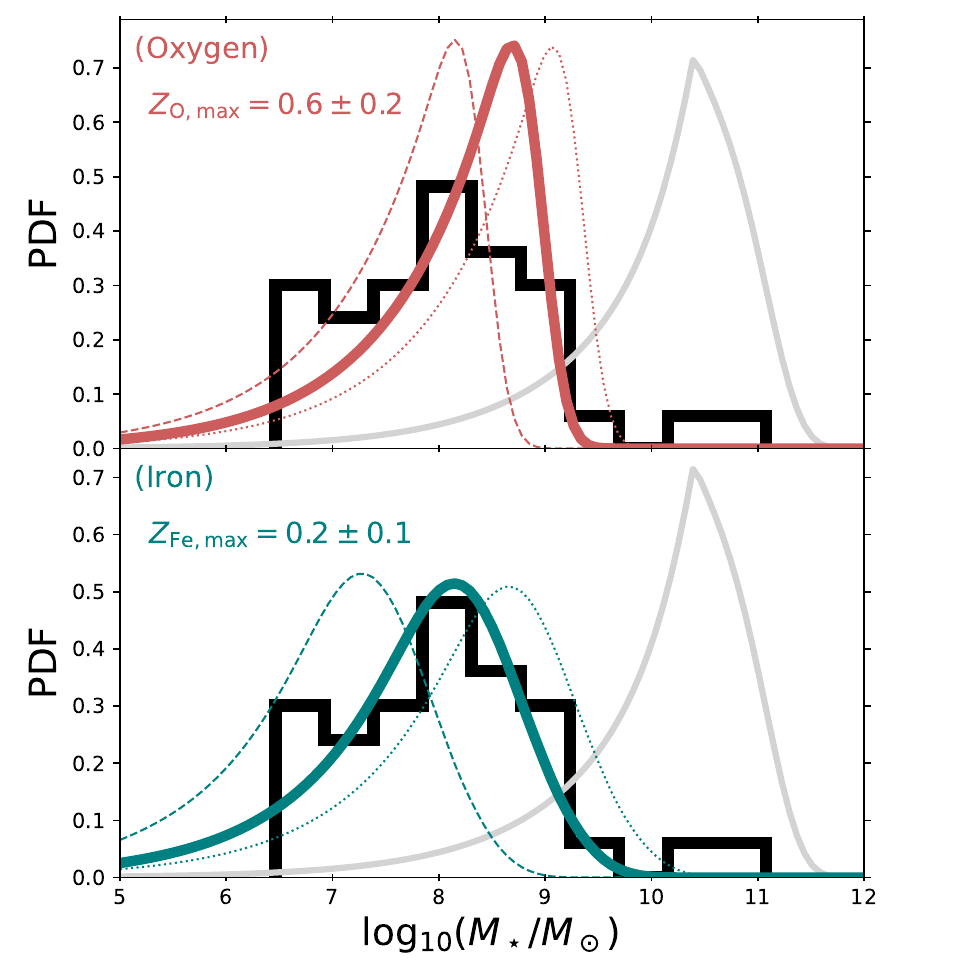}
    \caption{
    Distribution of SLSN host-galaxy stellar mass $\Mstar$ compared to model predictions. Black histograms show the PTF data from \cite{Schulze21}, comprising 37 SLSN hosts (with a median redshift of $z \approx 0.27$). Grey curves show the predicted distribution if SLSNe were unbiased tracers of star-formation. Colored curves instead show model predictions accounting for a maximum metallicity cutoff $Z_{\rm max}$ above which SLSN production is suppressed (Equation~\ref{eq:eta}).
    The bottom panel shows results assuming that iron elemental abundance sets the SLSN production rate (blue curves). The thick solid curve shows the preferred model, with $Z_{\rm Fe,max} = 0.2\,Z_\odot$, while thin dotted (dashed) curves show the variation if $Z_{\rm max}$ is increased (decreased) by $0.1$. 
    The top panel shows instead predictions for a scenario in which oxygen sets the SLSN production rate. This scenario is not physically-motivated, but it more closely follows previous approaches in the literature, so we show it here for comparison. The preferred model in this case has $Z_{\rm O,max} \sim 0.6\,Z_\odot$.
    }
    \label{fig:Zmax}
\end{figure}

The key parameter in the model described in \S\ref{sec:formalism} is the cutoff metallicity $Z_{\rm max}$ above which the production of SLSN is suppressed (Equation~\ref{eq:eta}). 
In previous work, this cutoff has been investigated by modeling the observed distribution of SLSN host-galaxy stellar mass. This distribution is truncated above $\Mstar \sim 10^9\, M_\odot$, well below the mass that would be expected if SLSNe were unbiased tracers of star-formation (See Figure \ref{fig:Zmax}). Because high-$M_\star$ galaxies have higher metallicity on average, this very observation originated the idea that SLSN might only occur in low-metallicity environments \citep[e.g.,][]{Neill11,Lunnan+14}

\cite{Schulze21} analyzed the host galaxy stellar-mass distribution of 37 SLSNe found by the Palomar Transient Factory (PTF) and found that this sample was best fit by a SFR-weighted stellar-mass function with an exponential truncation above $\log_{10}(\Mstar/M_\odot) = 8.64^{+0.46}_{-0.64}$. Using the \cite{mannucci2010} FMR, they relate this mass scale to an oxygen abundance $\OH_{\rm max} = 8.26^{+0.26}_{-0.30}$ that corresponds to a metallicity cutoff $Z_{\rm max} = 0.37^{+0.30}_{-0.19}~Z_\odot$.
Other studies have also found similar values for $Z_{\rm max}$ \citep[e.g.,][]{Perley16,Chen17a,schulze2018}. However, note that these values of $Z_{\rm max}$ are all determined assuming that oxygen is the relevant metallicity threshold.

In the following, we briefly repeat this exercise in order to obtain a rough estimate of $Z_{\rm max}$ that is applicable to our fiducial model, in which iron (and not oxygen) plays the main role in determining the SLSN metallicity threshold. One might naively assume that the same $Z_{\rm max}$ would apply to both oxygen and iron, however this is not necessarily true given that the conversion between oxygen to iron depends on galaxy sSFR (as a proxy for the recent star-formation history). 

We use the \cite{Schulze21} PTF data-set of SLSNe and compare the observed host-galaxy stellar-mass distribution to theoretical predictions based on different assumptions and different values of $Z_{\rm max}$.
Specifically, we calculate the predicted stellar-mass distribution at redshift $z$ as
\begin{equation}
    p( \Mstar \vert z ) \propto 
    \int \SFR \eta(Z) \varphi( \SFR \vert \Mstar, z ) \Phi(\Mstar \vert z) d\SFR 
    ,
\end{equation}
where $p(\Mstar)$ is normalized such that $\int p(\Mstar)d\Mstar=1$. This function depends on $Z_{\rm max}$ through Equation~(\ref{eq:eta}), as well as on the assumed element X that is relevant.
In the following calculations, we adopt a redshift of $z=0.2658$ corresponding to the median redshift of the \cite{Schulze21} SLSN sample.

Figure~\ref{fig:Zmax} shows the resulting probability distribution functions (PDFs) $p(\Mstar \vert z)$ under different sets of assumptions, in comparison to the SLSN host-galaxy data (black) from \cite{Schulze21}. Red curves shown in the top panel assume that the relevant metallicity threshold is set by the oxygen elemental abundance. The thick solid curve shows the expected distribution for $Z_{\rm O, max} = 0.6\,Z_\odot$ (equivalent to $\OH_{\rm max} = 8.47$), while thin dotted (dashed) curves show how this distribution evolves if this nominal value of $Z_{\rm O, max}$ is changed by $+0.2$ ($-0.2$). Although none of the curves fit the data particularly well, it is clear that $Z_{\rm 
O, max} \sim 0.6\,Z_\odot$ is preferred over higher (lower) values.
The bottom panel shows the same calculation but taking iron to be the key element of relevance. The thick solid blue curve shows the expected stellar-mass distribution for $Z_{\rm Fe,max} = 0.2\,Z_\odot$ (equivalent to $\FeH_{\rm max} = 6.80$), while thin dotted (dashed) curves show the resulting distributions if the value of $Z_{\rm Fe, max}$ was increased (decreased) by $0.1$. These results suggest that $Z_{\rm Fe, max} \sim 0.2\,Z_\odot$ is preferred by the data.
The plots also suggest that our physically-motivated Fe-dependent model better fits the data, when compared to the O-dependent model. Finally, we also show the expected stellar-mass distribution function if no metallicity cutoff is assumed. These are plotted as grey curves, which peak at $\log_{10}(\Mstar/M_\odot) \sim 10.5$. This highlights the importance of metallicity in governing SLSN production rates.

Our results for the oxygen-tracing model are broadly consistent with previous estimates of $Z_{\rm max}$ within the large uncertainties, although our inferred $Z_{\rm O, max} \sim 0.6\,Z_\odot$ for oxygen is on the higher end of other estimates \citep[][]{Perley16,Chen17a,schulze2018,Schulze21}. This may partly be due to different underlying assumptions about the stellar-mass function $\Phi(\Mstar)$ or the SFMS. Different papers have also used different scaling relations to convert galaxy mass to oxygen abundance (although \citealt{Schulze21} use the same FMR we adopt here). Importantly, we model the theoretical distributions $p(\Mstar)$ via a calculation that directly folds in metallicity. This is advantageous over previous approaches which typically model the mass distribution independently of metallicity, and translate the mass cutoff to a metallicity threshold using an MZR.

Our iron-tracing model (bottom panel) is both better motivated based on theoretical considerations and seems to fit the data better. The implied metallicity threshold $Z_{\rm Fe,max} \approx 0.2\,Z_\odot$ obtained from comparing this model to the stellar-mass distributions is somewhat lower than previously estimated values, but still broadly consistent within the uncertainties. 

It is important to note that we have not employed any formal fitting procedure in the current work. Partly, this is because such a procedure would also demand a treatment of selection effects. In particular, we expect that low-$M_\star$ galaxies would be under-represented in the data, given that these galaxies are less luminous and harder to detect. This may explain why our theoretical predictions (which do not correct for this bias) over-predict the distribution at low masses.

\subsection{SLSN Rate Evolution}
\label{sec:results}

\begin{figure}
    \centering
    \includegraphics[width=0.99\linewidth]{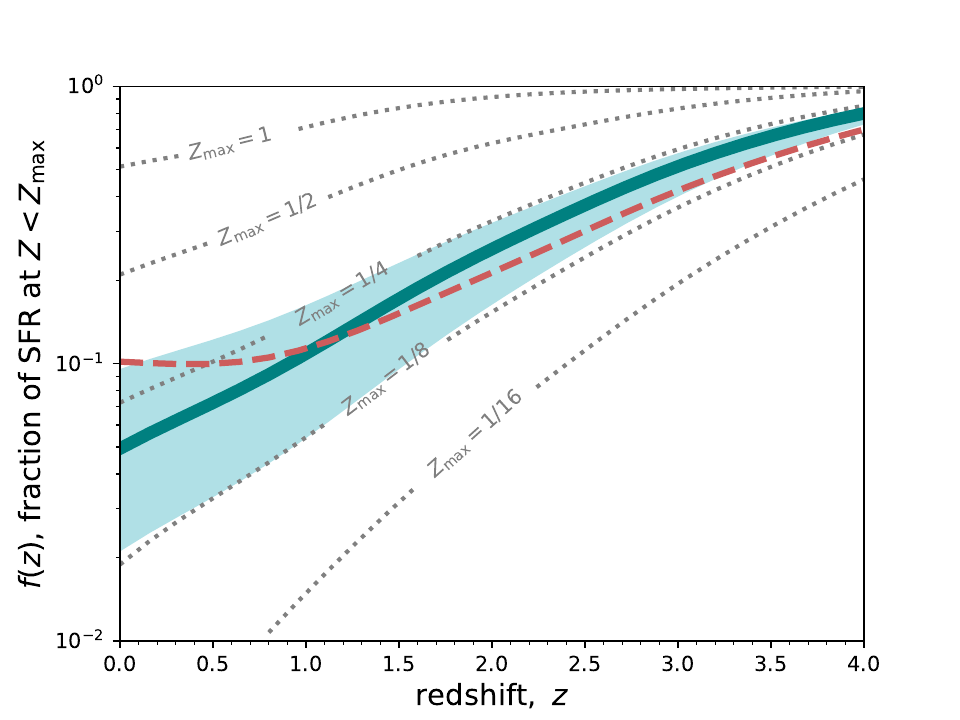}
    \caption{
    The function $f(z;Z_{\rm max})$ that underpins much of the calculations in our work (Equation~\ref{eq:f_final}). This function can be interpreted as the fraction of cosmic star formation at redshift $z$ that occurs in galaxies where the metallicity is lower than $Z_{\rm max}$.
    Dotted-gray contours show $f(z)$ for a range of $Z_{\rm Fe, max}$ (as labeled), assuming that SLSN-production tracks the abundance of iron and adopting the `mixed Fe-enrichment' oxygen-to-iron conversion model of \cite{Chruslinska+26}.
    Solid blue curve shows our fiducial SLSN model, which follows these assumptions and adopts a value of $Z_{\rm Fe,max}=0.2$ (\S\ref{sec:metallicity_cutoff}; Figure~\ref{fig:Zmax}). The light-blue shaded region shows the uncertainty associated with the O/Fe conversion prescription (quantified by taking the extreme cases of `slow' and `fast' Fe-enrichment models presented by \citealt{Chruslinska+26}, assuming a fixed $Z_{\rm Fe,max}=0.2$).
    Finally, the red dashed curve shows an alternative model whereby SLSN-production is governed by the abundance of O instead of Fe, taking $Z_{\rm O,max}=0.6$ motivated by \S\ref{sec:metallicity_cutoff}.
    }
    \label{fig:f_of_z}
\end{figure}

Using the formalism described above, we now proceed with a calculation of the function $f(z)$ and the associated SLSN rate following Equations~(\ref{eq:Rate},\ref{eq:f_final}).
Figure~\ref{fig:f_of_z} shows the function $f(z)$, which can be interpreted as the fraction of cosmic SFR at redshift $z$ that occurs below some threshold metallicity $Z_{\rm max}$. 
Dotted gray curves show $f(z)$ for different values of the maximum iron metallicity-cutoff $Z_{\rm max}$. As expected, the fraction of SFR that occurs at metallicity $Z<Z_{\rm max}$ decreases with lower values of $Z_{\rm max}$. Figure~\ref{fig:f_of_z} also shows that $f(z)$ generally increases at higher redshifts, reflecting the fact that average metallicity in the Universe decreases with redshift. This fact is at the crux of our current work: the relative number of low-metallicity massive stars born at high redshifts is larger than at $z \sim 0$, and therefore SLSNe are more frequent at higher redshifts than implied by simply scaling the local-Universe rates by the cosmic SFRD.

The solid blue curve in Figure~\ref{fig:f_of_z} shows our fiducial Fe-dependent model with $Z_{\rm Fe,max} = 0.2\,Z_\odot$ (see \S\ref{sec:metallicity_cutoff}).
This curve, similar to the dotted-gray curves, were calculated using the `mixed Fe-enrichment' model of \cite{Chruslinska+26} to convert between oxygen and iron abundances (Equation~\ref{eq:FeH}). There are large uncertainties associated with this O/Fe conversion, and these can appreciably affect the resulting $f(z)$ curves at fixed $Z_{\rm max}$. To illustrate this, the light-blue shaded region in Figure~\ref{fig:f_of_z} shows the range of $f(z;Z_{\rm Fe,max}=0.2)$ expected if instead we use the `slow Fe-enrichment' (upper boundary of shaded region) or `fast Fe-enrichment' models (lower boundary) from \cite{Chruslinska+26}. At redshift $z \sim 0$ this translates into a factor of $\sim 4$ uncertainty in the value of $f(z)$.
Finally, we also show an oxygen-dependent model whereby SLSNe occur only in galaxies whose oxygen abundance is below a metallicity of $Z_{\rm O,max} = 0.6\,Z_\odot$ (chosen based on the result of \S\ref{sec:metallicity_cutoff}). This model is illustrated with a dashed-red curve in Figure~\ref{fig:f_of_z}. Note that this curve cannot be directly compared to the dotted-gray contours because the latter are calculated based on the iron abundance. Consequently, the red dashed curve corresponding to $Z_{\rm O,max}=0.6\,Z_\odot$ falls significantly below the $Z_{\rm Fe,max} = 1/2$ dotted-gray curve.

For reference in future studies, we find that the numerically-calculated value of our fiducial Fe-tracing model is well-fit by the following fitting function,
$f(z; Z_{\rm Fe,max}=0.2) \approx 1 - \exp \left( -a e^{bz} \right)$
with $a \simeq 0.049$ and $b \simeq 0.89$. The maximum error of this fit is $\lesssim 3\%$. We stress that there is no physical motivation for this particular functional form, and in the remainder of this paper we use our numerically-calculated values for $f(z)$. However, we include this analytic fitting-function for completeness and convenience.

\begin{figure*}
    \centering
    \includegraphics[width=0.8\linewidth]{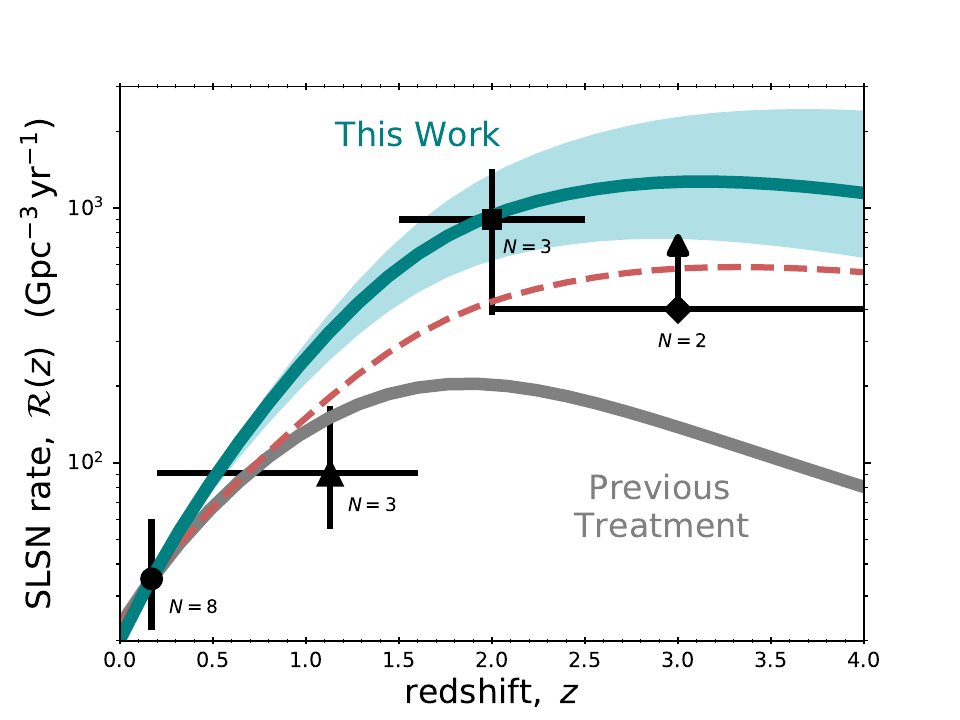}
    \caption{
    Volumetric rate of (Type-I) SLSNe as a function of redshift. Black points and their associated errorbars show measured rates and the number of SLSNe $N$ from which each rate is derived 
    (\citealt{Cooke+12}, diamond; \citealt{Prajs+17}, triangle; \citealt{Moriya+19}, square; \citealt{Frohmaier+21}, circle; see also \citealt{Quimby+13}). 
    The curves show different theoretical predictions for the rate evolution, all anchored to the measured rate at $z=0.17$ (\citealt{Frohmaier+21}; black circle).
    The gray curve shows the expected rate if SLSNe were unbiased tracers of star formation, as treated in most previous studies. In this case, the rate would evolve in lockstep with the cosmic star-formation rate density.
    The blue and red curves instead show the expected SLSN rate evolution when taking into account these transients' low-metallicity preference (solid blue: our fiducial model with Fe-dependence and $Z_{\rm max}=0.2$; dashed red: an O-dependent model with $Z_{\rm max}=0.6$). This is encapsulated via the function $f(z)$,
    which describes the fraction of star-forming galaxies that have sufficiently low metallicity to harbor SLSNe
    (Equation~\ref{eq:f_final}).
    Because galaxy metallicities are lower at higher redshifts, $f(z)$ is an increasing function of $z$, and the predicted high-redshift SLSN rate is significantly larger when accounting for this metallicity effect (see Figure~\ref{fig:f_of_z}).
    }
    \label{fig:rate}
\end{figure*}

Using the results described above, we now calculate the redshift evolution of the SLSN volumetric rate 
following Equation~(\ref{eq:Rate}). This implies that $\mathcal{R}(z) = \mathcal{R}(z_0) \Psi(z)f(z) / \Psi(z_0)f(z_0)$
where $\Psi(z)$ is the cosmic SFRD and $z_0$ is some reference redshift at which the SLSN rate $\mathcal{R}(z_0)$ is well-determined. We take $z_0 = 0.17$ and $\mathcal{R}(z_0) = 35 \, {\rm Gpc}^{-3}\,{\rm yr}^{-1}$ from \cite{Frohmaier+21}, and adopt $\Psi(z)$ from \cite{Madau&Dickinson14}. 
Our fiducial model assumes that iron governs the SLSN production efficiency and uses a maximum metallicity of $Z_{\rm Fe,max} = 0.2\,Z_\odot$, motivated by the results of \S\ref{sec:metallicity_cutoff} (see Figure~\ref{fig:Zmax}). The SLSN rate evolution implied by this fiducial model is shown by the blue curve in Figure~\ref{fig:rate}.
This can be compared to previous SLSN rate estimates in the literature, which typically ignore metallicity effects. This scenario assumes that the SLSN rate perfectly traces cosmic star formation (equivalent to setting $f=1$) and is shown by the gray curve in Figure~\ref{fig:rate}.
A comparison of these curves highlights the high-$z$ SLSN rate enhancement predicted by our formalism: our work suggests that the high-redshift SLSN rate may be $\gtrsim 10$ times larger than previously estimated.
Furthermore, the rate of SLSNe across cosmic time peaks at a redshift of $z \approx 3.3$, much higher than the redshift at which cosmic star formation peaks ($z \approx 1.9$). The maximum SLSN rate implied by our model is $\mathcal{R}(z = 3.3) \approx 1,300\,{\rm Gpc}^{-3}\,{\rm yr}^{-1}$, roughly six times higher than the maximum rate predicted by previous approaches (gray curve).
Note that these numbers are subject to significant uncertainty due to both uncertainties in the low-redshift SLSN rate chosen as an anchor (e.g., \citealt{Perley20,Frohmaier+21,Pessi+25b}) and systematic uncertainties associated with the different scaling relations adopted in our calculations (e.g., the choice of stellar-mass function, SFMS, and FMR; see \S\ref{sec:formalism} for further details). However, the qualitative finding of an enhanced SLSN rate at high redshifts holds irrespective of these uncertainties.

For comparison purposes, we also calculate the SLSN rate evolution for a model where the abundance of oxygen, not iron, sets the SLSN production efficiency. This model assumes $Z_{\rm O,max} = 0.6\,Z_\odot$ and is shown with a red curve in Figure~\ref{fig:rate}. To the best of our knowledge, there is no physical basis for this type of model. However, it can still be used as a reference point for our fiducial Fe-dependent model given the fewer assumptions that go into the oxygen-model calculation. 
Importantly, this model also predicts an enhanced SLSN rate at high redshifts compared to the baseline scenario in which SLSNe track the cosmic SFRD (grey curve).

We next compare our predicted SLSNe rates to existing observational constraints \citep[e.g.,][]{Cooke+12,Quimby+13,McCrum+15,Prajs+17,Moriya+19,Perley20,Frohmaier+21,Pessi+25b}.
Black points in Figure~\ref{fig:rate} show a subset of these constraints and their associated uncertainties (68\% confidence intervals on $\mathcal{R}$), focused on the higher-redshift population. The number of events $N$ from which each rate estimate is derived is also listed next to each of these points.
At the lowest redshift, the black circle shows the rate $35^{+25}_{-13}\,{\rm Gpc}^{-3}\,{\rm yr}^{-1}$ derived by \cite{Frohmaier+21} from $N=8$ SLSNe-I using PTF data. We use this rate as an anchor for our models, though note that this estimate is still based on a small number of events. Therefore, by construction, all of the curves coalesce at this point.
At increasing redshift, we also show: the rate derived by \cite{Prajs+17} based on $N=3$ SLSNe found in the Canada–France–Hawaii Telescope Supernova Legacy Survey (triangle); the rate from \cite{Moriya+19} based on $N=3$ SLSNe as part of the Subaru high-$z$ supernova campaign (square); the rate estimate by \cite{Cooke+12} based on $N=2$ SLSNe detected in the Canada-France-Hawaii Telescope Legacy Survey Deep Fields (diamond). 
We note that the latter two estimates do not distinguish between Type-I and Type-II SLSNe, so should be taken with caution. Furthermore, we plot the \cite{Cooke+12} rate as a lower-limit following the suggested interpretation of this constraint by the authors of that work.

Finally, we also note that some of the quoted rate measurements implicitly depend on the assumed rate evolution. For example, \cite{Prajs+17} derive their rate by simulating a population of events that have a fixed redshift-independent volumetric rate within the range $0.2 < z < 1.6$. They show that replacing this treatment with a simulated population whose rate instead tracks cosmic star-formation history does not affect the derived rate appreciably. Still, it is worth noting that the qualitative change was an increase in the derived rate, and that folding in metallicity evolution as implied by our current modeling would likely further increase the derived rate. A similar approach is also followed by \cite{Frohmaier+21}.

Interestingly, our fiducial model (blue curve) shows better agreement with the data compared to previous estimates (gray curve) even though it was not tuned to do so. Our fiducial model formally has only one free parameter, the value of the metallicity-cutoff $Z_{\rm max}$. The preferred value adopted in our fiducial model ($Z_{\rm Fe,max}=0.2\,Z_\odot$) was determined based on the observed host-galaxy mass distribution of low-redshift SLSNe (at $z \approx 0.3$). The fact that this value of $Z_{\rm max}$ can simultaneously explain the galaxy mass distribution (Figure~\ref{fig:Zmax}) and the high-redshift rate evolution (Figure~\ref{fig:rate}) is a non-trivial success of this model.
However, we caution again that there are significant systematic uncertainties associated with the modeling, so that our present results should be treated as a proof of concept.

\section{Discussion}
\label{sec:discussion}

The primary finding of our work is that integrating the expected metallicity dependence of SLSNe into rate calculations leads to a significantly enhanced SLSN rate at high redshifts (see Figure~\ref{fig:rate}).
In the following section, we discuss some implications of these findings.

First, the intrinsic rate of SLSNe $\mathcal{R}(z)$ directly impacts the detectability prospects of these events with wide-field deep surveys such as Vera C. Rubin's Legacy Survey of Space and Time (LSST) and Roman.
Previous studies estimating SLSN detection metrics with LSST assumed that the intrinsic SLSN rate evolves in lockstep with cosmic star formation \citep[][]{Villar+18,ELAsTiCC}. Our finding that the high-$z$ SLSN rate is enhanced compared to this naive approach implies that LSST will detect a significantly larger number of SLSNe than previously estimated.
We will present a detailed investigation of revised SLSN detection prospects with the Vera C. Rubin Observatory in a forthcoming companion paper. 

Using our present results, we can also estimate the SLSN production efficiency. In the CCSN literature such an efficiency is typically defined as the SN yield (number of SNe) per unit mass of stars formed,
$r_{\rm CCSN} = \mathcal{R}_{\rm CCSN}/\Psi$. \cite{Horiuchi11} find $r_{\rm CCSN} \approx 5 \times 10^{-3} /M_\odot$ using observational constraints from the Lick Observatory (\citealt{Li11}; see also \citealt{Pessi+25b}).
Because SLSN production appears strongly suppressed at high metallicity, we define an analogous efficiency parameter $\tilde{r}_{\rm SLSN}$ as the SLSN yield per unit mass of stars formed {\it with metallicity below $Z_{\rm max}$}. 
The SLSN efficiency is therefore 
$\tilde{r}_{\rm SLSN} = \mathcal{R}_{\rm SLSN}(z_0) / \Psi(z_0) f(z_0) \approx 3 \times 10^{-5} / M_\odot$ for our fiducial Fe-tracing model.\footnote{
Note that $\tilde{r}_{\rm SLSN}$ does not vary with redshift and therefore its value is independent of the choice of $z_0$.}
This implies that a single SLSN is produced for every $\sim 3 \times 10^4\,M_\odot$ of stars formed with metallicity $Z_{\rm Fe} < 0.2\,Z_\odot$. SLSNe are therefore a rare outcome even among the subset of high-mass stars that have low metallicity.

Another implication of our findings is that the relative rate of SLSNe to ``standard'' CCSNe will be enhanced at higher redshifts. At a low redshift of $z_0 = 0.17$, there are  an estimated $3500^{+2800}_{-720}$ CCSN events for every single SLSN \citep{Frohmaier+21}. 
Assuming that standard CCSNe track the cosmic SFRD (and e.g., are not affected by metallicity; though see \citealt{Pessi+23}),
then the relative rate of SLSNe at sufficiently high redshifts should be a factor $\sim 1 / f(z_0)$ larger than this estimate. For our fiducial Fe-tracing model $f(z_0) \approx 0.055$ which implies that there would be $\sim 200$ CCSNe for every single SLSN at sufficiently high redshifts $z \gtrsim 4$. 
This is also consistent with the calculations in the previous paragraph, which imply a relative rate of $\approx r_{\rm CCSN}/\tilde{r}_{\rm SLSN} \sim 200$ at high $z$.
A sufficiently enhanced rate of SLSNe compared to standard CCSNe implies that SLSNe could contribute appreciably to SN-driven galaxy feedback and nucleosynthesis at high redshifts, and must therefore be taken into account when modeling these processes. 
A rapidly-rotating magnetar---a leading candidate for the central engine powering SLSNe \citep{kasen10,Woosley10}---can inject up to $\sim10^{53}\,{\rm erg}$ of energy, roughly $\sim 100$ times more than a typical CCSN explosion \citep[e.g.][]{Metzger+15}. If this amount of energy were injected by each SLSN, then the time-averaged energy injection rate by SLSNe would approach that of normal CCSNe at high redshifts, and SLSNe would possibly have a comparable impact on galaxy-feedback as standard CCSNe.
However, light-curve modeling suggests that the typical energy injected by a single SLSN is only $\sim$several times larger than a standard CCSN \citep[e.g.,][]{Nicholl+17b,Chen+23}. Therefore, SLSNe likely remain subdominant on galaxy-feedback, at least for the rate-evolution implied by our current modeling.

\begin{figure}
    \centering
    \includegraphics[width=\columnwidth]{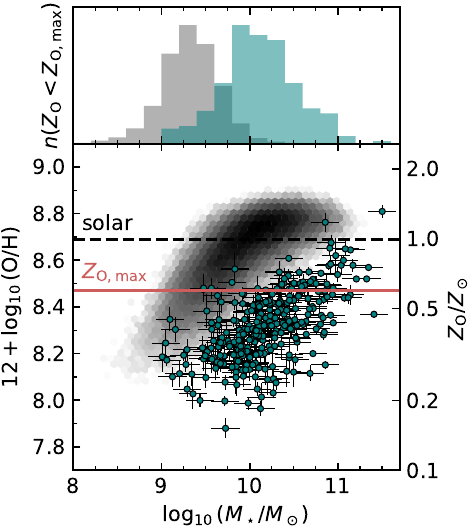}
    \caption{The correlation between gas-phase oxygen abundance (O/H) and stellar mass ($M_\star$) in galaxies at $z\sim0$ (grey histogram) and $z\sim2-3$ (blue points with error bars). Because galaxies have lower oxygen abundances at fixed $M_\star$ at higher redshifts, more massive galaxies will fall below the nominal SLSN suppression threshold at $Z_\textrm{O,max}=0.6$, corresponding to $12+\log_{10}(\textrm{O/H})=8.47$ (horizontal red line). The top panel shows the $M_\star$ distributions for local and Cosmic Noon galaxies with metallicities below this value, showing a shift of $\sim0.5$~dex. As more massive star-forming galaxies generally have higher SFRs, we should increasingly expect to see SLSN in more massive galaxies at higher redshifts.}
    \label{fig:mzr}
\end{figure}

Another implication of the evolving metallicity in the Universe is the expected properties of high-$z$ SLSN host galaxies. In the nearby Universe, SLSNe are found primarily in low-$M_\star$ dwarf galaxies. Indeed, this was the original evidence pointing towards these events' preference for low-metallicity environments \citep[e.g.,][]{Lunnan+14}.
However, this same effect implies that higher-$z$ SLSNe will occur in comparatively more massive galaxies, because the average metallicity at fixed $M_\star$ decreases with redshift \citep[e.g.,][]{erb2006,Sanders+21,stanton2026}. 
Figure~\ref{fig:mzr} shows the correlation between $M_\star$ and gas-phase oxygen abundance for local star-forming galaxies from the Sloan Digital Sky Survey \citep[SDSS DR8;][]{aihara2011}, alongside a representative sample of star-forming galaxies at Cosmic Noon from the Keck Baryonic Structure Survey \citep[KBSS;][]{steidel2014,strom2017}. The top panel shows the $M_\star$ distributions for galaxies at $z\sim0$ (grey) and $z\sim2$--$3$ (blue) with $12+\log_{10}({\rm O/H})<8.47$, which is the (oxygen) suppression threshold inferred using our analysis framework, illustrated by the horizontal red line. If metallicity is what drives the apparent preference for low-$M_\star$ hosts in the local Universe, then we should expect to see a larger number of high-$z$ SLSNe occurring in more massive galaxies, which contain more of the overall star formation at Cosmic Noon than their smaller counterparts. This will also be true if iron abundance is the underlying property that matters, as even comparatively massive star-forming galaxies are young and chemically immature enough to have O/Fe ratios consistent with CCSN yields \citep[e.g.,][]{,strom2022}; in other words, Fe/H will be $\sim3$--$4\times$ lower than O/H in the same $z\sim2$--$3$ galaxies, easily satisfying the lower $Z_{\rm Fe, max}$.
The distribution of SLSN host-galaxy properties as a function of redshift can be used as another diagnostic test of the metallicity dependence of SLSNe (e.g., as an additional means of constraining $Z_{\rm max}$ in Equation~\ref{eq:eta}).
Finally, we note that the higher mass of high-$z$ SLSNe host galaxies may be an important effect to consider for transient classifiers that rely on host-galaxy properties in their classification scheme \citep[e.g.,][]{Gagliano+21,Sheng+24}.

Our focus in this work has been on SLSNe, however, the formalism we have developed can similarly be applied to other classes of transients or astrophysical sources.
In particular, metallicity effects have been discussed in the context of long gamma-ray bursts (GRBs; e.g., \citealt{Fruchter+06,Savaglio+09,Svensson+10,Kruhler+15,Vergani+15,Japelj+16,Perley+16a,Palmerio+19}), stripped-envelope SNe (e.g., \citealt{Graur+17a,Graur+17b}), luminous fast blue optical transients (LFBOTs; \citealt{Perley+19b,Sevilla+26,Nugent+26b}), fast radio bursts (FRBs; e.g., \citealt{Sharma24_FRB}; though see \citealt{Horowicz&Margalit26})  and gravitational-wave events (e.g., \citealt{Belczynski+10,Dominik+13,Neijssel+19,Santoliquido+21,Briel22,Broekgaarden+22}).

For long GRBs, there has been some theoretical modeling in the literature with similarities to our present work \citep[e.g.,][]{Kocevski+09,Robertson&Ellis12,Ghirlanda&Salvaterra22}. 
To the best of our knowledge, however, our work is the first to apply these ideas in the context of SLSNe.
Furthermore, our formalism introduces a completely novel aspect---treating the metallicity effects of iron rather than oxygen.

Despite the advantages of our new formalism, there are many uncertainties associated with our analysis. Perhaps the largest uncertainty relates to the conversion between oxygen and iron abundances (Equation~\ref{eq:FeH}). The relative abundance of iron and oxygen is complicated because it depends on the detailed star-formation and assembly history of any given galaxy and will change over the course of a galaxy’s lifetime. In our current work, we have used the Fe/O conversion prescription recently derived by \cite{Chruslinska+24,Chruslinska+26}. These authors consider three different models, whose predictions vary appreciably from one another. We have adopted the `mixed Fe-enrichment’ model of \cite{Chruslinska+26} for our fiducial estimates, however, we have also briefly considered the uncertainty associated with this choice by instead adopting the `slow’ and `fast’ Fe-enrichment models as a bracket on the O/Fe relation. At a fixed iron-metallicity threshold $Z_{\rm Fe,max}$, these differing choices introduce appreciable systematic uncertainty in $f(z)$ and the associated high-redshift rate (see shaded blue bands in Figures~\ref{fig:f_of_z},\ref{fig:rate}). However, this comparison is not completely fair---a different choice of Fe/O model would require a different value of $Z_{\rm Fe,max}$ to fit the host-galaxy mass distribution (\S\ref{sec:metallicity_cutoff}; Figure~\ref{fig:Zmax}). Exploring this we find that the ‘slow Fe-enrichment’ model requires $Z_{\rm Fe,max} \approx 0.14\,Z_\odot$ while the `fast Fe-enrichment’ model requires $Z_{\rm Fe,max} \approx 0.32\,Z_\odot$ to match the host-galaxy mass distribution. 
Somewhat surprisingly, comparing these different models leads to $f(z)$ and rate $\mathcal{R}(z)$ predictions that are very close to our fiducial model. This suggests that our high-redshift rate predictions may be insensitive to the exact Fe/O conversion prescription, even though the implied values of $Z_{\rm Fe, max}$ are. However, we caution that this indicative test does not amount to an exhaustive investigation of this potential source of systematic uncertainty.
Furthermore, there are many additional sources of possible systematic error including the prescribed stellar-mass function $\Phi(M_\star,z)$, star forming main sequence, and FMR. Different choices for any of these relations can potentially affect the results.

While we incorporate scatter around the star-forming sequence, we have not treated scatter around the FMR in our current work. Future work may wish to extend the formalism to account for this, and investigate the sensitivity of the results to such scatter.
Finally, we have treated the possible effects of metallicity on SLSNe in a fairly simplistic manner. In our present formalism, we have assumed that metallicity only plays a role in modifying the production efficiency of SLSNe (with respect to star formation). 
The functional form we have chosen for this metallicity dependence (Equation~\ref{eq:eta}) is ad hoc, and the value of $\beta$ in this equation is set by hand. These assumptions could be relaxed to explore more general metallicity dependencies if there is physical motivation to do so, or if the data can distinguish between such possibilities.
Furthermore, one might be interested in extending our formalism to allow metallicity to play a role in the SLSN luminosity function
(although the apparent absence of redshift evolution found by \citealt{Hsu+21} tentatively points against this). 
While this is outside the scope of our current work, we encourage future work to investigate these issues.

\section{Conclusions}
\label{sec:conclusions}

In this work, we have investigated the high-redshift rate evolution of SLSNe
taking into account these events' preference for low metallicity environments. 
Our primary findings can be summarized as follows:
\begin{itemize}
    \item We have developed a novel framework through which metallicity dependence on the birth rate of astrophysical sources can be treated (\S\ref{sec:formalism}). Importantly, this framework distinguishes between the metallicity traced by oxygen abundance and iron-group element metallicity. This is important because the two do not evolve in lockstep, and because iron likely plays a more dominant role in determining the progenitor evolution and formation efficiency of sources.
    \item Applying our formalism to SLSNe, we find that the host-galaxy mass distribution of low-redshift SLSNe ($z \sim 0.3$) can be reproduced if SLSNe  only occur in environments where the iron metallicity is lower than $Z_{\rm Fe,max} \approx 0.2\,Z_\odot$ (\S\ref{sec:metallicity_cutoff}; Figure~\ref{fig:Zmax}).
    \item Using this estimated value of $Z_{\rm Fe,max}$, we calculated the predicted SLSN rate evolution $\mathcal{R}(z)$ and find that the predicted rate exceeds previous estimates---which ignored metallicity effects---by a factor of $\gtrsim 10$ at high redshifts (Figure~\ref{fig:rate}).
    This is a consequence of the fact that a larger fraction of cosmic star formation occurs in low-metallicity environments at higher redshifts, because the average metallicity in the Universe decreases with $z$.
    \item 
    Our fiducial model implies a peak volumetric rate of $\mathcal{R} \approx 1{,}300\,{\rm Gpc}\,{\rm yr}^{-1}$ at a redshift of $z \approx 3.3$. This is a much higher rate than predicted by naive estimates which assume that SLSNe are unbiased tracers of star formation. Correspondingly, the SLSN rate peaks at higher redshifts than the cosmic star-formation ($z \approx 2$).
    \item These new rate estimates better align with existing observational constraints on the high-redshift rate of SLSNe (Figure~\ref{fig:rate}) and help explain the order-of-magnitude discrepancy first noted by \cite{Cooke+12}.
    However, these constraints are based on very small samples and are subject to large uncertainties.
    The Vera C. Rubin Observatory is set to transform this situation by detecting large numbers of SLSNe at high redshifts \cite[e.g.,][]{Villar+18}. 
    In upcoming work we will present revised estimates for SLSN detection prospects by Rubin. 
\end{itemize}

\begin{acknowledgments}
This work was made possible through Scialog grant \#SA-LSST-2024-094 from Research Corporation for Science Advancement with support from Heising-Simons Foundation.
B.M. is supported in part by the National Science Foundation under grant number AST-2508620. A.L.S is supported by the David and Lucile Packard Foundation (Packard Fellowship, grant 2024-77399) and the National Science Foundation (grant number AST-2406780).
\end{acknowledgments}

\bibliography{sample701, papers}{}
\bibliographystyle{aasjournal}



\end{document}